\documentstyle[11pt]{article}
\begin{document}
\newcommand{\be}{\begin{equation}}
\newcommand{\ee}{\end{equation}}
\newcommand{\bea}{\begin{eqnarray}}
\newcommand{\eea}{\end{eqnarray}}
\newcommand{\beaa}{\begin{eqnarray*}}
\newcommand{\eeaa}{\end{eqnarray*}}
\newcommand{\qd}{\quad}
\newcommand{\qqd}{\qquad}
\newcommand{\npb}{\nopagebreak[1]}
\newcommand{\nn}{\nonumber}
\newcommand{\prel}{\Preliminaries}
\newcommand{\theor}{Theorem}
\newcommand{\defi}{Definition}
\newcommand{\prop}{Proposition}
\newcommand{\rem}{Remark}
\newcommand{\example}{Example}
\title{\bf  Finsler-Geometrical Approach to the Studying of Nonlinear
Dynamical Systems}
\author{V.S. Dryuma\thanks{Work supported in part by MURST,Italy.
Permanent address: Institute of Mathematics
 Academy of Sciences of Moldova
Kishinev, MD 2028, Moldova; Academitcheskaya str.5, e-mail:
15valery@mathem.moldova.su, valery@gala.moldova.su}
\hspace{0.1em} and M.Matsumoto\\
15 Zenbu-cho, Shimogamo, Sakyo-ku, Kyoto, Japan, 606}
\date{}
\maketitle
\begin{abstract}
A two dimensional Finsler space associated with the differential equation
$y''=Y_3 y'^3+Y_2 y'^2+Y_1 y'+Y_0$ is characterized by a tensor equation and
called the Douglas space. An application to the Lorenz nonlinear dynamical
equation is discussed from the standpoint of Finsler geometry.
\end{abstract}
\section{Introduction}

    The differential equation is usually the most appropriate mathematical
tool for analyzing a dynamical system. In the 1960's E. Lorenz used a
computer to model weather patterns, using a set of ordinary nonlinear
equations
$$
\dot x= k(y - x),\quad \dot y=rx-y-xz,\quad \dot z=xy- bz.
$$
This is perhaps the most celebrated set of nonlinear ordinary differential
equations.

    Since 1992 the first author has derived the second order differential
equation y'' = f(x,y,y') from the Lorenz set and continued to study it from
the standpoint of the differential geometry ( [4] , [5] , [6]). Two of his
results was very attractive to the second author:

      1) The Lorenz equation
coincides with the differential equation of geodesics of a two-dimensional
space which belongs to the special class of Finsler spaces, called the Berwald
(affinely connected) spaces [4], and

      2) The necessary condition (presented in invarianten under point
transformations form) for determination of
the Finsler metrics for equations type $y''=Y_3 y'^3+Y_2 y'^2+Y_1 y'+Y_0$, [5].

     We remember various geometrical investigations of the differential
equation of the form
$$
y'' = Y_3(x,y)y'^3 + Y_2(x,y)y'^2 + Y_1(x,y)y' + Y_(x,y),
$$
(e.g., [3]). The Lorenz equation is just of this form. Recently the second
author and S.Bacso [2] have succeeded in the characterization by a tensor
equation of Finsler spaces whose geodesic equation is of this form; those
spaces are called Douglas spaces. Therefore the remarkable results of the
first author can be described as an interesting theory based on the Finsler-
geometrical foundations.

\section{Preliminaries}

     We consider an n-dimensional Finsler space $F^n = (M^n, L(x,y))$ on a smooth
n-manifold $M^n$ ([1], [8]). The fundamental function $L(x,y)$, a real-valued
function on the tangent bundle $TM^n$ , is usually supposed certain conditions
from the geometrical standpoint, but only the homogeneity and the regularity
are mainly important for our following considerations.

     1. $L(x,y)$ be positively homogeneous in $y^i$ of degree one:
$$
L(x,py) =  p L(x,y), for \vee p > 0.
$$

     2.
$L(x,y)$ be regular:
$$
g_{ij}=\dot \partial_{i}\dot \partial_{j} F
$$
 has non-zero $g = \det g_{ij}$, where $F = L^2/2$ and $\dot\partial_{i}
=\frac{\partial}{\partial y_i}$.

     Let $(g^{ij})$ be the inverse matrix of $(g_{ij})$ and construct
$$
2\gamma^i_{jk}(x,y)= g^{ir}(\partial_k g_{rj}+\partial_j g_{rk}-
\partial_r g_{kj}),
$$
$$
2G^{i}(x,y) = g^{ij}\{(\dot \partial_j  \partial_r F)y^r-\partial_j F\},
$$
where $\partial_i=\frac{\partial}{\partial x^i}$. Then we have
$\gamma^i_{jk}(x,y) y^j y^k=2G^i(x,y)$.

    The length $s$  of an arc C: $x^i = x (t)$ on $M^n$ is defined by the
integral $s = \int L(x,\dot x)dt$, $\dot x^i =\frac{dx^i}{dt}$, and the
extremely of the integral, called the geodesic, is given by the Euler
differential equations $d(\dot \partial L)/dt-\partial_i L=0$, which are
written in the form
\begin{equation}
\dot x^i\{\ddot x^j+2G(x,\dot x)\}-\dot x^j\{\ddot x^i+2G(x,\dot x)\}=0.
\label{geo}
\end{equation}

In order to introduce geometrical quantities in $F^n$, we are concerned with
a Finsler connection $F\Gamma = (F^i_{jk}(x,y), N^i_j(x,y), V^i_{jk}(x,y))$
on $F^n$. For a tensor field $F\Gamma$ gives rise to the h and v - covariant
differentiations: We treat of a tensor field $X^i (x.y)$ of (1,0)-type, for
brevity. Then we get two tensor fields as follows:
$$
\bigtriangledown^h_j X^i=\delta_j X^i+ X^r F^i_{rj}(x,y),
$$
$$
\bigtriangledown^v_j X^i=\dot\partial_j X^i+ X^r V^i_{rj}(x,y),
$$
where $\delta_j=\partial_j-N^r_j(x,y)\dot \partial_r$. The h and v-covariant
derivatives $\bigtriangledown^h X$ and $\bigtriangledown^v X$
are tensor fields of (1,1)-type.

    In the following we needs the Berwald connection $B\Gamma =
    (G^i_{jk}(x,y), G^i_j(x,y), 0)$, where
$$
G^i_j=\dot \partial_j G^i, \quad G^i_{jk}=\dot \partial_k G^i_j.
$$
$"V^i_{jk} = 0"$ for $B\Gamma$ means that $\bigtriangledown^v_j =\dot
\partial_i$ in $B\Gamma$. We shall denote by (;) the h-covariant
differentiation $\bigtriangledown^h$ in $B\Gamma$. Then we obtain the
commutation formulae, called the Ricci identities:
\begin{eqnarray}
X^i_{;j;k}-X^i_{;k;j}=X^r H^i_{rjk}-X^i_{;r}R^r_{jk},\\ \nonumber
\dot\partial_k(X^i_{;j})-(\dot \partial_k X^i)_{;j}=X^rG^i_{rjk}.
\end{eqnarray}\label{eq}

    $R^i{jk}$ is called the (v)h-torsion tensor, defined by
$$
R^i_{jk}=\partial_k G^i_j-G^i_{jr}G^r_k-[j,k],
$$
where $[j,k]$ denotes the interchange of indices j,k of the preceding terms.
$H^h_{ijk}$ and $G^h_{ijk}$ are called the h and hv- curvature tensors
respectively, defined by
\begin{equation}
H^h_{ijk}=\dot\partial_i R^h_{jk},\quad G^h_{ijk}=\dot\partial_i G^h_{jk}.
\end{equation}\label {def}
   It is noted that $G^h_{ijk}$ is symmetric in the subscripts. The contracted
tensor  $G^r_{rjk}=G_{jk}$ called the hv-Ricci tensor.

     For the later use we shall give the three classes of special Finsler
spaces as follows:

      1. Riemannian spaces, characterized by $g_{ij}=g_{ij}(x)$, that is ,
the C-tensor $C_{ijk}=(\dot\partial_k g_{ij})/2$ vanishes.

      2. Locally Minkowski spaces, characterized by the existence of the
adapted coordinate system $(x^i)$ such that $L = L(y)$. Then $G^i_{jk}= 0$
and ~(\ref{geo}) is reduced to $\dot x^i\ddot x^j-\dot x^j\ddot x^i=0$. The
tensorial characterization is $"R^i_{jk}=0$ and $G^h_{ijk}=O".$

      3. Berwald spaces, characterized by $G^i_{jk}=G^i_{jk}(x)$, that is,
$G^h_{ijk}= 0$.

     The classes (1) and (2) are contained in the class (3).

     We consider a change of Finsler metric: $F^n = (M^n, L(x,y))$
 $\rightarrow$ $\bar F^n= (M^n,\bar L(x,y))$. If any geodesic of $F^n$ coincides
 with a geodesic of $\bar F^n$ as a set of points and vice versa, then the
 change is called projective and $F^n$ is said to be projectively related to
 $\bar F^n$ ( [1],[7]).
$F^n$ is projectively related to$\bar F^n$, if and only if there exists a
scalar field P(x,y), positively homogeneous in $y^i$ of degree one,
satisfying
$$
\bar G^i(x.y) = G^i(x,y) + P(x,y)y^i.
$$
If we put $P_i=\dot\partial_i P$, then we get
$$
\bar G^i_j=G^i_j+P_jy^i+P\delta^i_j.
$$

From these relations we obtain the invariant of protective change as follows:
\begin{equation}
Q^h = G^h-\frac{1}{n+1}G^r_ry^h.\label {inv}
\end{equation}

    Consequently we are led to the following projective invariants by means
of successive differentiation with respect to $y^i$:
$$
Q^h_i=\dot\partial_i Q^h=G^h_i-\frac{1}{n+1}(G^r_{ri}y^h+G^r_r\delta^h_i),
$$
$$
Q^h_{ij}=\dot\partial_j Q^h_i=G^h_{ij}-\frac{1}{n+1}(G_{ij}y^h+G^r_{ri}
\delta^h_j+G^r_{rj}\delta^h_i),
$$
where $G_{ij}$ is the hv-Ricci tensor. Further we get the Douglas tensor
\begin{equation}
D^h_{ijk}=\dot\partial_k Q^h_{ij}=G^h_{ijk}-\frac{1}{n+1}
\{(\dot\partial_k G_{ij})y^h+G_{ij}\delta^h_k+G_{kj}\delta^h_i+
G_{ik}\delta^h_j\},\label{doug}
\end{equation}
where $G^h_{ijk}$ is the hv-curvature tensor.

      A Finsler space $F^n= (M^n, L(x,y))$ is said to be with rectilinear
extremals, if $M^n$ is covered by coordinate neighborhoods in which any
geodesic is represented by n linear equations $x^i = x^i_0 + ta^i$ in a
parameter t, where $x^i_0$ and $a^i$ are constants.

     Next a Finsler space is called projectively flat, if it has a
covering by coordinate neighborhoods in which it is projectively related to
a locally Minkowski space. A Finsler space is projectively flat, if and only
if it is with rectilinear extremals. We have the well-known theorem as
follows:

\begin{theor} Pf.

     A Finsler space of dimension n is projectively flat, if and only if

    1. $n \geq 3$: $W^i_{jk}=0$ and $D^h_{ijk}=0$,

    2. n=2: $K_{ij}=0$ and $D^h_{ijk}=0$,

here $D^h_{ijk}$ is the Douglas tensor. $"W^i_{jk}=0"$ is equivalent to the
fact that $F^n$ be of scalar curvature. On the other hand, $"K_{ij}=0"$ for
$n = 2$ is a differential equation satisfied by the h-scalar curvature R
(or the Gauss curvature, cf. p. 4).
\end{theor}

\section{ Douglas spaces}

    The present section is devoted to the short introduction to the recent
theory which was given by [2].

     We shall start our discussions from the equations ~(\ref{geo}) of
geodesics of a two-dimensional Finsler space $F^2$. If we denote
$(x^1, x^2)$ by (x.y), take x as the parameter t and use the symbols
$y' = dy/dx$, $y'' = dy'/dx$, then ~(\ref{geo}) (i=1, j=2) for $F^2$ is
written in the form
\begin{equation}
y" = f(x,y,y') = X_3 y'^3 + X_2 y'^2 + X_1 y' + X_o, \label{geo1}
\end{equation}
where $X_3 = G^1_{22}$, $X_2 = 2G^1_{12} - G^2_{22}$, $X_1 =G^1_{11}-2G^2_{12}$,
$X_0=-G^2_{11}$ and $G^i_{jk}=G^i_{jk}(x,y,1,y')$ [1O].

     If we are specially concerned with a Riemannian space $F^2$, then
$G^i_{jk}=\gamma^i_{jk}$ are usual Christoffel symbols, and hence $X's$ of
~(\ref{geo1}) do not contain $y'$. Next, if $F^2$ is a Berwald space, then
$G^i_{jk}$ do not contain y' by definition. Consequently f(x,y,y') of those
spaces is a polynomial in y' of degree at most three.

     This special property of f(x,y,y') is equivalent to the fact that $
\dot x^1 G^2(x,\dot x)-\dot x^2G^1(x,\dot x)$ of ~(\ref{geo}) is a homogeneous
polynomial in $\dot x^1,\dot x^2$ of degree three.

Generalizing this fact, we shall give

\begin{defi}

   A Finsler space $F^n$ is said to be of Douglas type
or called a Douglas space, if $D^{ij}(x,y)=G^i(x,y) y^j-G^j(x,y) y^i$ are
homogeneous polynomials in $y^i$ of degree three.
\end{defi}

\begin{prop} 1.

     A Berwald space is of Douglas type, where
$G^i(x,y)$ of ~(\ref{geo}) are of the form $G^i_{jk}(x)y^jy^k/2$
\end{prop}

\begin{theor} 1.

    A Finsler space $F^2$ off dimension two is of Douglas
type, if and only if, in every local coordinate system (x,y) the
differential equation $y''=f (x,y,y')$ of geodesics is such that
f(x,y,y') is a polynomial in y' of degree at most three.
\end{theor}

\example\ 1.

  ([13] , [14]). We consider a Randers space $R^2$  of
dimension two, that is, the metric being $L(x,y)=\alpha+\beta$, $\alpha^2=
a_{11}(x,y)\dot x^2+2a_{12}(x,y)\dot x \dot y+a_{22}(x,y)\dot y^2$,
$\beta=b_{1}(x,y)\dot x+b_{2}(x,y)\dot y$.
Suppose that the Riemannian $\alpha$ be positive-definite, and hence we can refer to
an isothermal coordinate system $(x.y)$ such that
$$
\alpha=a E, \quad a=a(x,y) > 0, \quad E=\sqrt {\dot x^2+ \dot y^2}.
$$

     Then the equation of geodesics of $R^2$ is written in the form
$$
ay'' + (a_x y' - a_y)(1 + y'^2) = (b_{1y} - b_{2x})(1 + y'^2)^{3/2}.
$$

Consequently $R^2$ is not of Douglas type in general; $R^2$ is of
Douglas type, if and only if $b_{1y}-b_{2x}=0$.

On the other hand, it is shown that a Kropina space of dimension two,
whose metric is $L = \alpha^2/\beta$, is a Douglas space.

      We treat of $D^{lm}=G^l y^m-G^m y^l$ in the Definition. $D^{lm}$ are
homogeneous polynomials in $y^i$ of degree three, if and only if
$D^{lm}_{hijk}=\dot\partial_k\dot\partial_j\dot\partial_i\dot\partial_h
D^{lm}=0$. We have the relations between $D^{lm}_{hijk}$ and
the Douglas tensor $D^h_{ijk}$ as follows:
\begin{eqnarray}
D^{lr}_{hijr}=(n+1)D^l_{hij}, \label{ten}\\
D^{lm}_{hijk}=(\dot \partial_k D^l_{hij})y^m+D^l_{ijk}\delta^m_h+
D^l_{jkh}\delta^m_i+D^l_{khi}\delta^m_j+D^l_{hij}\delta^m_k-[l,m].
\end{eqnarray}
Consequently, if $D^{lm}_{hijk}=0$, then the first relation implies
$D^l_{hij}=0$, and if $D^l_{hij}=0$, then the second relation implies
$D^{lm}_{hijk}=0$. Therefore we have Fundamental

\begin{theor}

     A Finsler space is of Douglas type, if
and only if the Douglas tensor $D^h_{ijk}$ vanishes identically.
\end{theor}

     Thus the special property of geodesics of a two-dimensional Finsler
space, stated in Theorem 1 , has been characterized by the tensor equation
$D^h_{ijk}=0$ in the viewpoint of Finsler geometry. Since
$D^h_{ijk}=\dot\partial_k\dot\partial_j\dot\partial_i Q^h$ from ~(\ref{doug}),
we have

\begin{theor} 2.

    A Finsler space is of Douglas type, if and only if $Q^h(x,y)$ of
~(\ref{inv}) are homogenious polynomials in $y^i$ of deqree two.
\end{theor}

     Thus, for a Douglas space $F^n$, we can put
$$
G^h=\frac{1}{n+1}G^r_r y^h +\frac{1}{2}Q^h_{ij}(x)y^iy^j,
$$
which shows that ~(\ref{geo}) can be written in the form
\begin{equation}
\dot x^i\ddot x^j -\dot x^j \ddot x^i = \{Q^i_{hk}\dot x^j- Q^j_{hk}\dot x^i\}
\dot x^h \dot x^k.\label{geo3}
\end{equation}

      In the two-dimensional case ~(\ref{geo3}) may be written as
\begin{eqnarray}
y'' = Y_3 y'^3 + Y_2 y'^2 + Y_1 y' + Y_0 ,\nonumber \\
Y_3=Q^1_{22},\quad Y_2=2Q^1_{12}-Q^2_{22},\quad Y_1=Q^1_{11}-2Q^2_{12},
\quad Y_0=-Q^2_{11},\label{geo4}
\end{eqnarray}
where $Q^i_{jk}=Q^i_{jk}(x,y)$ do not contain y'.

\section{Two-dimensional Douglas space}

      The present section is devoted to studying Douglas spaces of dimension
two. Let $F^2 = (\pi (x,y), L(x,y;p,q)$ be a two-dimensional Finsler space,
which is defined on the (x,y) plane $\pi(x,y)$ and has the fundamental
function L(x,y;p,q). Since this L is positively homogeneous in (p,q) of degree
one, we can introduce
$$
W=\frac{L_{pp}}{q^2}=-\frac{L_{pq}}{qp}=\frac{L_{qq}}{p^2},
$$
called the Weierstrass invariant ([10],[12]). Then the Euler
equation $d(\dot\partial_i L)/dt-\partial_i L=0$ of geodesic can be written in
the single equation
\begin{equation}
p\dot q - \dot p q+\frac{1}{W}(L_{xq}- L_{yp}) = 0, \label {Eul}
\end{equation}
and ~(\ref{geo}) shows $\frac{1}{W}(L_{xq}- L_{yp})=2(p G^2-q G^1)$.
Therefore we have

\begin{theor} 3.

    A two-dimensional Finsler space is a Douglas space,
if and only if $\frac{1}{W}(L_{xq}- L_{yp})$ is a homogeneous polynomial in
 (p.q) of degree three.
\end{theor}

\example \ 2.

([11] , (3.7b)). We deal with a two-dimensional Finsler space $F^2$ with the
metric
$$
L = (q-p)\log|z-1| - (q+p)log|z+1|- 2xq,\quad z=\frac{q}{p}.
$$
We have $\frac{1}{W}(L_{xq}- L_{yp})= p(p^2 - q^2)$. Thus $F^2$ is a Douglas
space. The differential equation of geodesics is $y'' = y'^2 - 1$.

     For $F^2= (\pi(x.y), L(x,y:p,q))$ we introduce the associated
fundamental function A (x, y, z) of three arguments by A(x,y,y')=
L(x,y;1,z). Then we have the relation between L and A :
\begin{equation}
L (x,y;p,q) = p A (x,y,\frac{q}{p}). \label{lag}
\end{equation}

     If we put $A'=\partial A/\partial z$, then we get
\begin{equation}
L_{xq}- L_{yp}=zA'_y+A'_x-A_y,\quad W=\frac{A''}{p^3}\label{lag1}
\end{equation}

Consequently ~(\ref{Eul}) is written in the form
\begin{equation}
A''y'' + y' A'_y + A'_x - A_y=0,\quad z=y',\label{ra}
\end{equation}
which is called the Rashevsky form [4].

     Now we consider the equation ~(\ref{geo1}) of a geodesic.
From ~(\ref{ra}) it follows that
\begin{equation}
A''f + z A'_y + A'_x - A_y=0,\quad z=y',\label{ra1}
\end{equation}
must be identically satisfies by (x,y,z) [5] ., where f = f(x,y,z)
and A = A(x,y,z). Differentiate ~(\ref{ra1}) successively by z: Putting
$S = \log\mid A''\mid$ and $P = S_x + zS_y$, we obtain
\begin{equation}
S'f + f' + P=0,\label{ra2}
\end{equation}
\begin{equation}
S''f + S'f' + f'' + P'=0,\label{ra3}
\end{equation}
\begin{equation}
S'''f + 2S''f' + S'f'' + f'''+P''=0,\label{ra4}
\end{equation}
and
\begin{equation}
S^{1V}f + 3S'''f' + 3 S''f'' +S'f'''+f^{1V}+ P'''=0.\label{ra5}
\end{equation}

     Suppose that $F^2$ be a Douglas space. Then $f^{1V} = 0$ from Theorem 1.,
and hence ~(\ref{ra5}) is reduced to
\begin{equation}
 S^{1V}f + 3S'''f' + 3 S''f'' +S'f'''+ P'''=0.\label{ra6}
\end{equation}

    Then the coefficients of $(f, f', f'', f''', 1)$ in the above five 
equations of (3.5) must satisfy 
$$
\Delta (A)=\left |\begin{array}{ccccc}
A''& 0 & 0 & 0 & \delta (A) \\
S'& 1 & 0 & 0 & P \\
S''& S' & 1 & 0 & P' \\
S'''& 2S'' & S' & 1 & P'' \\
S^{1V}& 3S''' & 3S'' & S' & P'''
\end{array} \right |=0,
$$
where $\delta (A)=z A'_{y}+A'_{x}-A_{y}$.

\begin{theor} 4

     A two-dimensional Finsler space is a Douglas space,
if and only if the associated fundamental function $A(x,y,z)$
satisfies $\Delta (A) = 0$, where $S = \log|A''|$, $P = S_x +zS_y$ and
$\delta(A)=zA'_y+A'_x-A_y$.
\end{theor}

     Proof: Only the sufficiency must be shown. From $\Delta (A)=0$ it
follows that the five linear equations
\begin{eqnarray}
A''x_1 + (zA'_y+ A'_x - A_y)x_5=0,\\
S'x_1 + x_2 + Px_5=0, \\
S''x_1+ S'x_2 + x_3 + P'x_5=0, \\
S'''x_1 + 2S''x_2 + S'x_3 + x_4 + P''x_5=0,\\
S^{1V}x_1 + 3S'''x_2 + 3S''x_3 + S'x_4 + P'''x_5=0,
\end{eqnarray}
has a non-trivial solution $(x_1,\cdots, x_5)$. Suppose that $x_5=0$.
Then the first relation gives $x_1 = 0$ because of $A''=pW \not= 0$ from
~(\ref{lag1}). Hence the second leads to $x_2=0$, the third to $x_3=0$ and
fourth to $x_4=0$,  which is a contradiction. Thus we have non-zero $x_5$.
Hence ~(\ref{geo}) and ~(\ref{ra1}) lead
to $f =x_1/x_5$. Then the second, comparing with ~(\ref{ra2}),
gives $f'=x_2/x_5$. Similarly we obtain $f''=x_3/x_5$
and $f'''=x_4/x_5$. Consequently fifth gives ~(\ref{ra6}), and,
comparing with ~(\ref{ra5}), $f^{1V}=0$ is concluded. Therefore f(x,y,z)
is a polynomial in z of degree three.

\begin{rem}

1) The determinant  given in the previous papers ([5], [6]) for the
differential equation such as ~(\ref{geo4}) is necessary, but not sufficient.
It must be corrected to $\Delta (A) = 0$ as above.

(2) According to the Fundamental Theorem and Theorem 4, it is sure that
$\Delta (A) = 0$ should coincide with vanishing of the Douglas
tensor in the two- dimensional case. In fact, $\Delta (A)$ is constructed
from A(x,y,z) by the differentiation one time with respect to (x.y) and six
times with respect to z. The Douglas tensor is the set of components
$D^h_{ijk}$, constructed from $L(x^i,y^i)$ in the same way, that is, by the
differentiation one time with respect to $x^i$ and six times with respect
to $y^i$.
\end{rem}

\example\ 3

([12], (4.12)). We treat of $F^2$ with
$$
L = 2p \log|\frac{q}{p}|+qu(x,y),
$$
where u(x,y) is a function of (x,y). From ~(\ref{lag}) we have
$A(x,y,z) = 2 \log|z| + zu$. Hence $S = \log 2 - 2 \log|z|$, $P = 0$ and it
is easy to show $\Delta (A) = 0$ for any u(x,y). Thus $F^2$ is a Douglas
space. The geodesic equation is given by $2y'' = u_x y'^2$.

     The differential equation $\Delta (A) = 0$, which is expanded in the
form
\begin{eqnarray}
A''[S'''_x+ zS'''_y + 3S''_y - \{3S''- S'^2\}(S'_x+zS'_y + S_y)-\nonumber\\
- S'(S''_x+zS''_y+2S'_y)- (S_x+ zS_y)\{3S''' - 5S'S'' + S'^3\}]-\\
\delta (A)\{S^{1V} - 4S'S''' - 3S''^2 + 6S'^2 S'' - S'^4\} = 0,\nonumber
\end{eqnarray}
has fundamental importance on the Inverse Problem of the Calculus of
Variations for ODE.

     There are many types of reductions of this equation to differential
equations of functions which are obtained from A(x,y,z) by lessening the
number of variables.

    A remarkable example of such reduction corresponds to the choice of
 A(x,y,z) in the form ([6], (8))
$$
A(x,y,z) = x^{(\beta/\alpha)-2}\omega(\xi,\eta),\quad \xi=y^{\alpha}
x^{-\beta},\quad
\eta=z^{\alpha}x^{\alpha-\beta}.
$$
Then the initial equation is reduced to the differential equation
of $\omega(\xi,\eta)$. Further the reduction
$$
A(x,y,z) =x^{\alpha}y^{-\alpha-1}z^2\omega(\xi), \quad \xi=\frac{zx}{y},
$$
leads to the ordinary differential equation of $\omega(\xi)$ ([5] , (40)).

     Let us consider an example of solution. Let function $A(x,y,z)$ be in
form
$$
A=\frac{\omega(z)}{x},\quad A_{zz}=\frac{\omega^{''}}{x}.
$$
Then we get
$$
S=\log A_{zz}=\log\omega^{''}-\log x,
$$
and
$$
A_{zx}=-\frac{\omega^{'}}{x^2},\quad S_z=\frac{\omega^{'''}}{\omega^{''}},
\quad S_x=-\frac{1}{x}.
$$

      The equation take the form
$$
A_{zz}[-S_x(S_z^3+3S_{zzz}-5S_zS_{zz})] +
A_{zx}[S_z^{4}+4S_zS_{zzz}-6S_z^2 S_{zz}-S_{zzzz}+3S_{zz}^2]=0,
$$
or
$$
\omega''[S'^3+3S'''-5S'S''] -
\omega'[S'^{4}+4S'S'''-6S'^2 S''-S^{1V}+3S''^2]=0.
$$

     After calculation all derivatives we obtain the equation
$$
3\omega^{V}-14\frac{\omega''' \omega^{1V}}{\omega''}+
12\frac{\omega'''^{3}}{\omega''}=
\omega' \left [24\frac{\omega'''^{4}}{\omega''^{4}}+8\frac{\omega''' \omega^{V}}
{\omega''^{2}}-36\frac{\omega'''^{2} \omega^{1V}}{\omega''^{3}}-
\frac{\omega^{V1}}{\omega''}+6\frac{(\omega^{1V})^2}{\omega''^{2}}\right].
$$

      This equation may be solved by the means of simple transformations.

      Using the substitution
$$
\omega''=T(\omega')=T(Q),
$$
we get the equation
$$
T'^3T+3T'''T^3-2T^2T'T''=
$$
$$
Q(T'^4-3TT'^2T''+T^2T'T'''-T^3T^{1V}+
2T^2T''^2).
$$
If we let
$$
T'=R(Q)T(Q), \quad T''=(R'+R^2)T, \quad T'''=(R''+3RR'+R^3)T,\quad
$$
$$
T^{1V}=(R'''+3R'^2+4RR''+6R^2R'+R^4)T,
$$
we find that function $R$ satisfies the equation
$$
3R''+7RR'+2R^3+Q(R'''+3RR''+R'^2+2R^2 R')=0.
$$

     Then using the substitution
$$
R=\frac{1}{Q}U(\log Q),
$$
we find the equation
$$
U'''+3(U-1)U''+U'^2 +2(U-1)^2U'=0,
$$
or after the change of variable $Z=U-1$
$$
Z'''+3ZZ''+Z'^2+2Z^2Z'=0.
$$

      For solution of this equation we present  the function $Z$ in form
$$
Z'=Y(Z)
$$
from which is followed the equation
$$
YY''+Y'^2+3ZY'+Y+2Z^2=0.
$$
It has particular solutions
$$
Y=-\frac{1}{2}Z^2,\quad Y=-\frac{3}{2}Z^2,
$$
and in general case can be reduce to the Abel's type of equation
using the substitutions.
$$
Y=Z^2V(\log Z).
$$
and
$$
V'=W(V).
$$

     Hence we get
$$
WW'+\frac{1}{V}W^2+(\frac{3}{V}+7)W+6V+7+\frac{2}{V}=0.
$$
Or
$$
\chi\chi'+(3+7V)\chi+6V^3+7V^2+2V=0,
$$
where
$$
W=\frac{\chi(V)}{V}.
$$

     The particular solution
$$
Y=-\frac{1}{2}Z^2
$$
lead to the function $\omega$ in the form
$$
\omega'=A\exp[{-\frac{1}{Bz+C}}],
$$
where $A,\ B,\ C$ are constants. It is corresponded the equation
$$
y''=\frac{1}{Bx}(By'+C)^2.
$$

\begin{rem}

    From the above equations we get the function $f(x,y,z)$ in form
$$
f = \frac{P'''-S'P''+(S'^2-3S'')P'-(S{'}^{3}+3 S'''-5S'S'')P }
 { S'^4+4S'S'''-6S'^2 S''+3S{''}^{2}-S^{1V} }.
$$
and corresponding expressions for its derivatives with respect to z
$f'$, $f''$ and $f'''$.
     So, for determination of the Finsler metric for a given equation
$y''=a_1(x,)y'^3+a_2(x,y)y'^2+a_3(x,y)y'+a_4(x,y)$ we must solve
corresponding system of nonlinear equations.
\end{rem}
\section{Geodesics of 1-form metrics}

     A Finsler metric L(x,y) of dimension n is called a 1-form metric,
if there is a standard Minkowski metric $L(v^{\alpha})$, where
$\alpha=1,\cdots,n$, in a real vector n-space $V^n$ with a fixed base and
$L(x,y) =L(a^{\alpha})$, where $a^{\alpha}=a^{\alpha}_i(x)y^i$ are n 1-forms
in $y^i$. These $a^{\alpha}$ must be independent;
$d = det(a^{\alpha}_i) \not=0$.

     Let $b^i_{alpha}$ be the inverse matrix of
$(a^{\alpha}_i)$ and put
\begin{equation}
F^i_{jk}(x)=b^i_{\alpha}\partial_k a^{\alpha}_j. \label{fin}
\end{equation}

     These give rise to the linear connection $(F^i_{jk}(x)$ ([1], 1.5.2.)
and
\begin{equation}
a^{\alpha}_{j;k}=\partial_j a^{\alpha}_i-a^{\alpha}_r F^r_{jk}=0.\label{fin1}
\end{equation}
Thus $a^{\alpha}$, $\alpha=1,\cdots,n$, are n covariant constant vector
fields.

     We introduce the Finsler connection $F1 = (F^i_{jk}, F^i_{0j}, 0)$,
$F^i_{0j}=y^rF^i_{rj}$ called the 1-form connection. Putting $L_{\alpha}=
\partial L/\partial a^{\alpha}$, we get $L_{(j)} (=\dot\partial_j L)=
L_{\alpha}a^{\alpha}_j$ and ~(\ref{fin1}) gives $L_{(j);i}=0$, which
implies $L_{i(j)}= L_{(j)(r)} F^r_{0i}+L_{r}F^r_{ji}$, $L_i=\partial_i L$.
Hence
$$
L_{i(j)} - L_{(j)i} = L_{(j)(r)} F^r_{0i} - L_{(i)(r)} F^r_{0j} -
L_{(r)} T^r_{ij},
$$
where $T^r_{ij}= F^r_{ij}-F^r_{ji}$ is the (h)h-torsion tensor of F1.

    Now we consider a two-dimensional Finsler space $F^2$ with 1-form
metric $L(a^1,a^2)$. Then
$$
M = L_{1(2)} - L_{2(1)} = L_{(2)(r)}F^r_{01} -L_{(1)(r)}F^r_{02}-L_{\alpha}
T^{\alpha},
$$
where $T^{\alpha}= T^i_{12} a^{\alpha}_i$.

     Analogously to the Weierstrass invariant W, we can define from
$L_{\alpha \beta}$, $\alpha, \beta = 1, 2$,
$$
w=\frac{L_{11}}{(a^2)^2}=-\frac{L_{12}}{a^1 a^2}=\frac{L_{22}}{(a^1)^2},
$$
called the intrinsic Weierstrass invariant of $F^2$. It is easy to
show from $L_{(i)(j)}=L_{\alpha\beta} a^{\alpha}_i a^{\beta}_j$
\begin{equation}
W=wd^2.\label{w}
\end{equation}

    Then $M = W(pF^2_{00}- qF^1_{00}) - L_{\alpha}T^{\alpha}$ and ~(\ref{Eul})
 gives the equation of geodesics in the form
\begin{equation}
p\dot q - \dot p q + p F^2_{00}- q F^1_{00}-
\frac{1}{d^2w}L_{\alpha}T^{\alpha}=0. \label{eul1}
\end{equation}

\begin{prop}\ 2.

    The equation of geodesics of a two-dimensional Finsler space with 1-form
metric $L(a^1,a^2)$ is written as ~(\ref{eul1}), where $F^i_{jk}$ are
defined by ~(\ref{fin}), $d = det(a^{\alpha}_i)$, "w" is the intrinsic
Weierstrass invariant and $^{\alpha}=(F^i_{12}-F^i_{21})a^{\alpha}_i$.
\end{prop}

     Since $F^i_{00}=F^i_{jk}(x)y^j y^k$ are polynomials in $y^i$ of degree
two and d does not contain $y^i$, ~(\ref{eul1}) leads to

\begin{theor}\ 5.

     A two-dimensional Finsler space with 1-form metric is a Douglas space,
if and only if $L_{\alpha}T^{\alpha}/w$ is a homogeneous polynomial
 in $y^i=(p,q)$ of degree three.
\end{theor}

     We shall use the following symbols for $(a^1, a^2)$ for brevity,
$$
a^1=a_1p+a_2q,\quad a^2=b_1p+b_2q,
$$
$$
a_{ik}=\partial_k a_i,
\quad b_{ik}=\partial_k b_i,
$$
where $a_i$ and $b_i$ are functions of (x,y).Then the connection
coefficients $F^i_{jk}$ are written as
\begin{equation}
F^i_{jk}=-\frac{1}{d}(a_2b_{jk}-b_2a_{jk}),\quad
F^2_{jk}=\frac{1}{d}(a_1b_{jk}-b_1a_{jk}).\label{con}
\end{equation}

    Now we are concerned with Berwald spaces, specially simple
Douglas spaces. In the two-dimensional case we refer to the Berwald frame
(1,m) in order to discuss such spaces ([1], 3.5). Then the (v)h-torsion tensor R* and the C-tenser Ci,* are written in the form
$R^{i}_{jk}$ and C-tensor $C_{ijk}$ are written in the form
$$
R^i_{jk}=\epsilon L R m^i(l_j m_k-j_k m_j), \quad LC_{ijk}=I m_i m_j m_k,
$$
where $\epsilon=\pm 1$ is the signature; $g_{ij}=l_il_j+\epsilon m_i m_j$.
The scalar R
and I are called the h-scalar curvature (or the Gauss curvature)
and the main scalar respectively.

All the Berwald spaces of dimension two are divided into three classes as
follows:

(1) R = 0 and $I \not = const$,

(2) R = 0 and $I = const$,

(3) R =$\not=0$ and $I = const$.

A Berwald space belonging to the class (1) or (2) is a locally Minkowski
space, and hence its geodesics is written as $y'' = 0$ in an adapted coordinate
system (x.y).

     We shall deal with two-dimensional Berwald spaces with the constant main
 scalar I. All of them has the 1-form metric $L(a^1,a^2)$ and are divided into
 three classes, according as the signature $\epsilon$ and the main scalar I
 as follows:
$$
B (1): \epsilon = +1,\quad I^2=4\frac{J^2}{(1 + J^2)} < 4,\quad
 L=\sqrt {(a^1)^2 + (a^2)^2} \exp (J \arctan \frac{a^2}{a^1}),
$$
$$
B (2): \epsilon = +1,\quad
I^2=4, \quad L=a^1 \exp \frac{a^2}{a^1},
$$
$$
B(3): L = (a^1)^{p}(a^2)^{1-p},\quad p \not = 0, 1,
$$
$$
(a)\quad  \epsilon = +1, \quad p<0 or p>1,\quad I^2=(2p-1)^2/p(p-1) > 4,
$$
$$
(b) \quad \epsilon = -1, \quad  0<p<1,\quad I^2=(2p-1)^2/p(1-p).
$$

     $L_{\alpha}T^{\alpha}/w$ of B(1), B(2) and B(3), appearing in
~(\ref{eul1}) respectively given as follows:
$$
B(1): \frac{1}{w}L_{\alpha}T^{\alpha}=\frac{(a^1)^2+(a^2)^2}{1+J^2}
     [(a^1-Ja^2)T^1+(Ja^1+a^2)T^2],
$$
$$
B(2): \frac{1}{w}L_{\alpha}T^{\alpha}=(a^1)^2[(a^1-a^2)T^1+a^1T^2],
$$
\begin{equation}
B(3): \frac{1}{w}L_{\alpha}T^{\alpha}=-a^1a^2[\frac{a^2}{1-p}T^1+
     \frac{a^1}{p}T^2].\label{Ber}
\end{equation}

\section{Lorenz dynamical system and Finsler metrics}

    Lorenz's nonlinear dynamical system is given by
$$
\dot x= k(y - x),\quad \dot y=rx-y-xz,\quad \dot z=xy- bz,
$$
where k, b are positive constant and r is a parameter. This is equivalent to
the following second order differential equations:
$$
k(y - x)y'' + k y'^2- \frac{1}{x}[ky- (b + 1)x]y'- \frac{y}{x}+\frac{1}{k(y-x)}
[x^2 y + b(y - rx)] =0.
$$
If we transform (x,y) to $(u = x, v = 1/(y - x))$ and write (u,v) as (x,y)
again, then the above is rewritten as ([4], [5], [6])
\begin{equation}
y''=\frac{3}{y}y'^2 +(\frac{1}{x}-my)y'+(nx^3-lx)y^4+(nx^2+t)y^3-
\frac{s}{x}y^2, \label{lor}
\end{equation}
where we put
\begin{equation}
m=1+\frac{b+1}{k},\quad n=\frac{1}{k^2},\quad l=\frac{b(r-1)}{k^2},\quad
t=\frac{b(k+1)}{k^2},\quad s=1+\frac{1}{k}.\label{par}
\end{equation}

The purpose of the present section is to find the two-dimensional Finsler
spaces $F^2$ whose geodesics are give by ~(\ref{lor}), To do so,we shall
pay attention to Berwald spaces $F^2$ belonging to the class
B(3):
$$
L = (a^1)^{p}(a^2)^{1-p},\quad p \not = 0, 1, \
$$
$$
a^1=a_1 x+a_2 y,\quad a^2=b_1 x+b_2 y,
$$
where $a_i$ and $b^i$ $i = 1,2$, are functions of (x,y). The equation of
geodesics is written from ~(\ref{eul1}, ~(\ref{con}) and ~(\label{Ber}) as
$$
y'' = Y_3 y'^3 + Y_2 y'^2 + Y_1 y' + Y_0 ,
$$
where we put
\begin{eqnarray*}
Y_3=F^1_{22}-\frac{1}{d^2}(\frac{b_2}{1-p}T^1+\frac{a^2}{p}T^2)a_2b_2,\\
Y_0=-F^2_{11}-\frac{1}{d^2}(\frac{b_1}{1-p}T^1+\frac{a^1}{p}T^2)a_1b_1,
\end{eqnarray*}
\begin{equation}
Y_2=F^1_{12}+F^1_{21}-F^2_{22}-\frac{1}{d^2}[\frac{b_2}{p-1}(a_1b_2+
2a_2b_1)T^1+\frac{a^2}{p}(a_2b_1+2a_1b_2)T^2], \label{co}
\end{equation}
$$
Y_1=F^1_{11}-F^2_{12}-F^2_{21}-\frac{1}{d^2}[\frac{b_1}{p-1}(a_2b_1+
2a_1b_2)T^1+\frac{a^1}{p}(a_1b_2+2a_2b_1)T^2].
$$

    Since ~(\ref{lor}) has no term containing $y'^3$, we first pay attention to
$Y_3$: ~(\ref{con}) gives
$$
Y_3=-\frac{1}{d}(a_2 b_{2y}-b_2 a_{2y})-\frac{1}{d^2}(\frac{b_2}{1-p}T^1+
\frac{a_2}{p}T^2)a_2 b_2.
$$
It is observed that $a_2 = 0$ or $b_2 = 0$ implies $Y_3 = 0$ at once.
Consequently, assume that $a_2 = 0$ in the following, and hence $a_1=1$
be assumed according to the special form of $L$. Thus we put
\begin{equation}
a_1=\dot x,\quad a^2 = b(x,y)\dot x + a(x,y)\dot y. \label{coef}
\end{equation}

Consequently we have from ~(\ref{co}) and ~(\ref{con})
\begin{eqnarray}
Y_2=-\frac{a_y}{a},\quad Y_1=\frac{1}{pa}[(1-p)a_x-(1+p)b_y], \\
Y_0=\frac{1}{pa^2}[b(a_x-b_y)-ab_x].\label {ko}
\end{eqnarray}
Consequently, comparing ~(\ref{lor}) with ~(\ref{ko}), our problem is to
find two functions a(x,y) and b(x,y) which satisfy
\begin{eqnarray}
\frac{a_y}{a}=-\frac{3}{y},\\
\frac{1}{pa}[(1-p)a_x-(1+p)b_y]=\frac{1}{x}-my,\label{lorc}\\
\frac{1}{pa^2}[b(a_x- b_y)-a b_x]=(nx^3-lx)y^4+(nx^2+t)y^3-\frac{s}{x}y^2.
\end{eqnarray}
Suppose that $p\not=-1$. Then above relations lead to a(x,y) and b(x,y) as
follows:
\begin{eqnarray}
a(x,y) =\frac{f(x)}{y^3},\label{a,b}\\
b(x,y) =\frac{p-1}{2(p+1)}\frac{f'(x)}y^2+\frac{p}{p+1}(\frac{1}{2xy}-m)
\frac{f(x}{y}+ g(x),
\end{eqnarray}
where f(x) and g(x) are functions of x alone. Substituting ~(\ref{a,b}) in
~(\ref{lorc}), we obtain the equation of the form $c_4y^4 + c_3y^3 + c_2y^2
+ c_1y = 0$, where c's are functions of x alone. Since $m\not=0$ from
~(\ref{par}), c's = 0 give the following four equations:
\begin{equation}
g=\frac{p+1}{m}(1-nx^2)xf,\label{del}
\end{equation}
$$
(p+1)xg'-(1+2x\frac{f'}{f})g=[\frac{m^2p}{p+1}-(p+1)(nx^2+t)]xf,\\
$$
\begin{equation}
(2p-1)(p-1)mxf'+[2(p+1)^2s-3mp]f=0,\label{me}
\end{equation}
\begin{equation}
(p^2-1)f''=(p-1)f'(\frac{2f'}{f}-\frac{p-1}{x})+p(p+2)\frac{f}{x^2}.\label{me1}
\end{equation}
    It is observed that g(x) is given from f(x) by the firsts of these
relations is rewritten in the form
\begin{equation}
(p-1)(nx^2 -1)xf'+ [(3p+2-m)nx^2+\frac{pm^3}{(p+1)^2}-pl-mt]f=0. \label{Per}
\end{equation}
Thus we shall find f(x) to satisfy ~(\ref{Per}), and previous relations.
They are of the form $A_if' + B_if = 0$,\ $i = 1,2$.

    From$f\not=0$ it follows that $A_1B_2-A_2B_1=0$ should be satisfied:
$$
(2p-1)(p-1)mx[(3p+2-m)nx^2+\frac{pm^3}{(p+1)^2}-pl-mt]-
$$
$$
(p-1)(nx^2-l)x[2(p+1)^2s-3pm]=0.
$$

   Dividing by $(p-1)x$, this gives the equation on form $Cx^2+D=0$,
which implies $C=D=0$:
\begin{equation}
2(p+1)^2s-3pm-(2p-1)(3p+2-m)m=0,\label{ver}
\end{equation}
\begin{equation}
(2p-1)[\frac{pm^3}{(p+1)^2}-pl-mt+(3p+2-m)l]=0.\label{ver1}
\end{equation}
We shall return to ~(\ref{me}). Assume that $2p-1\not=0$. Then ~(\ref{ver})
enables us to define
\begin{equation}
q=\frac{2(p+1)^2s-3mp}{(2p-1)(p-1)m}=\frac{3p+2-m}{p-1},\label{huy}
\end{equation}
and ~(\ref{me}) is written as
\begin{equation}
f'=-\frac{q}{x}f,\label{muz}
\end{equation}
which implies $f''=(q+1)qf/x^2$. Hence ~(\ref{me1}) is now reduced to
$$
[(p-1)q-p][(p-1)q+p+2]=0.
$$
therefore we divide our discussion into the following two cases:
$$
(i) \quad q=\frac{p}{p+1},\quad  (ii) \quad q=-\frac{p+2}{p-1}.
$$
We deal with the case (i).

     ~(\ref{huy}) and ~(\ref{ver1}) give $m=2(p+1)$,
$s=2p$, $t=4p$. Hence ~(\ref{par}) leads to $k=1/(2p-1)$, $b=2/(2p-1)$,
$n=(2p-1)^2$, $l=2(r-1)(2p-1)$. Consequently ~(\ref{muz}) and ~(\ref{del})
give f and g respectively. Thus a(x,y) and b(x,y) are given by ~(\ref{a,b})
as
\begin{eqnarray}
a(x,y)=c x^{\frac{p}{1-p}}y^{-3},\label{ha} \\
b(x,y)=cx^{\frac{1}{1-p}}[(2p-1)(r-1)-\frac{1}{2}(2p-1)^2x^2-\frac{2p}{xy}],
\end{eqnarray}
where c is a non-zero constant. Since k and b are positive, we have
$p > 1/2$.

     The case (ii) is analogously treated and we obtain $k=1/(3-4p)$,
      $b = 8p/(3-4p)$, $r=(4p-1)/(3-4p)$. Since k and b are positive, we
have $0 < p < 3/4$, $p\not=1/2$.
\begin{eqnarray}
a(x,y)=c x^{\frac{p+2}{p-1}}y^{-3}, \label{kl}\\
b(x,y)=cx^{\frac{3}{p-1}}[8p(2p-1)x^2-\frac{(3-4p)^2}{4}x^4+
\frac{1}{y^2}-\frac{4px}{y}],
\end{eqnarray}

where c is a non-zero constant.

     Further we consider the case p = 1/2. Then ~(\ref{ver}) gives m = 3s
and ~(\ref{me1}) is reduced to
\begin{equation}
3x^2ff'' - 4(xf')^2 - xff' + 5f^2=0. \label{do}
\end{equation}
~(\ref{Per}) gives f' = fh(x), where
\begin{equation}
h(x)=\frac{ux^2-v}{(nx^2-l)x},\quad u(x) =(7-2m)n,\quad v=l+2mt-
\frac{4}{9}m^3.
\label{re}
\end{equation}
Then ~(ref{do}) is rewritten as $3x^2h' - (xh)^2-xh+5=0$. Substituting from
~(\ref{re}), this is written in the second order equation in $x^2$. Equating the
three coefficients to zero, we obtain
$$
(u-n) (u+5n) = 0, \quad (u+5n) (v-l)=0, \quad (v-l)(v+5l)=0.
$$
From u = n it follows that m = 3 from ~(\ref{re}) , and hence s = 1 , which
contradicts to ~(\ref{par}). Thus the above leads to the following two cases:

(i) u =-5n, \quad v=l, \quad (ii)\quad u=-5n,\quad v=-5l.

    We treat of the case (i). Then ~(\ref{re}) gives m = 6, t = 8, s = 2,
and hence ~(\ref{par}) yields k = 1, b = 4, n = 1, l = 4(r-1). ~(\ref{re})
gives $h(x)=-[5x^2 + 4(r-1)]/[x^2 - 4(r-1)] x$ and f' = fh yields $f=cx/E^3$,
$E = x^2- 4 (r-1)$. Therefore we obtain
\begin{eqnarray}
a(x,y)=\frac{cx}{(Ey)^3},\quad E=x^2-4(r-1), \label{sol}\\
b(x,y) = \frac{cx}{E^2}[\frac{x}{(Ey)^2}-\frac{2}{Ey}-\frac{x}{4}].
\end{eqnarray}

    Next, it is easy to show that the case (ii ) leads only to a
special case where r = 1 of (i).

    Finally we deal with the case p = -1. Since ~(\ref{lorc}) give
$a = f(x)/y^3$ and $2a_x/a=my - 1/x$ respectively, we have
$my - 1/x = 2f' (x)/f(x)$, which implies m = 0, which contradicts to
~(\ref{par}).
Summarizing up all the above, we have

\begin{theor}\ 6

      The Lorenz equation ~(\ref{lor}) with ~(\ref{par}) is regarded
as the equation of geodesics of the two-dimensional Berwald space with the
fundamental function $L (x,y;\dot x,\dot y) =\dot x^p (b\dot x + a\dot y)^{1-p}$, if k, b
and r satisfy the following conditions, where a(x,y) and b(x.y) are
given as follows:

    (1)\ k = 1/(2p-1), b = 2/(2p-1), r = arbitrary, where $p > 1/2$,
        p $\not = 1$,

        a(x,y) and b(x,y) are given by ~(\ref{ha}).

    (2) k = 1/(3-4p), b = 8p/(3-4p), r = (4p-1)/(3-4p), where $0 < p < 3/4$,

        p $\not=1/2$. a (x, y) and b (x, y) are given by ~(\ref{kl}).

    (3) k = 1 , b = 4, r = arbitrary, where p = 1/2. a(x,y) and b(x,y)

     are given by ~(\ref{sol}).

\end{theor}

\begin{rem}\ 1

    The metric $L =\sqrt {\dot x(b\dot x+a\dot y)}$ of the case (3) is a
Riemannian metric with the signature $\epsilon=-1$.
\end{rem}

\section*{Conclusion}

    The values of parameters plays a crucial role in
geometry of equation. At the change of parameters the behaviour of its
integral curves (and, correspondengly, its Geometry!) my be change radically.
The Finsler-Geometrical approach to studying nonlinear dynamical systems
the equivalent to ODE's of type $y''=Y_3 y'^3+Y_2 y'^2+Y_1 y'+Y_0$  makes
possible to investigate all kinds of Geometries
connected with such type of equations. Fundamental partial differential
equation $\Delta=0$ for the Finsler metrics is basis of this approach and it
gives the hope to understand the nature of chaos from geometrical point
of view. It will be the object of next work.

\section*{Acknowledgement}
This work has been supported partial by grant INTAS-93-0166 and the first
author is very grateful to Physical Department of the Lecce University
(Italy) for financial support, which allowed him to continue his scientific
activity during the last years.

\section{References}
\begin{itemize}
\item[1] {P.L.Antonelli, R.S.Ingarden and M.Matsumoto:
The theory of Sprays and Finsler Spaces with Applications in Physics
and Biology, Kluwer Acad. Publishers, Dordrecht, 1993.}
\item[2]{S.Bacso and M.Matsumoto: Publ- Math. Debrecen 51 (1997), 385.}
\item[3]{E.Cartan: Lecons sur la theorie des espaces a connexion
projective, Gauthier-Villars, 1937.}
\item[4]{V.S.Dryuma: Rep. of 1X Internl. Conf. on Topology and its
Applications, 12-16, Oct., 1992, Ukraine, Kiev, p.70.}
\item[5]{V.S.Dryuma: Theor. Mat. Fiz. 99 (1994), 241.}
\item[6]{V.S.Dryuma: Proceedings of the First  Workshop on
Nonlinear Physics, 1995, p.83.}
\item[7]{M.Matsumoto: Tensor, N.S. 34 (1980), 303.}
\item[8]{M.Matsumoto: Foundations of Finsler Geometry and Special Finsler Spaces,
Kaiseisha Press, Otsu, Japan, 1986.}
\item[9]{M.Matsumoto: J.Math. Kyoto Univ. 29 (1989), 489.}
\item[10]{M.Matsumoto: Open syst. and Inform. Dynamics 3 (1995), 291.}
\item[11]{M.Matsumoto: J. Math. Kyoto Univ. 35 (1995), 357.}
\item[12]{M.Matsumoto: Tensor, N.S., to appear.}
\item[14]{M.Matsumoto and H.-S.Park: Rev. Roum. Math., to appear.}
\end{itemize}

\end{document}